\documentclass{ifacconf}

\usepackage{graphicx}      
\usepackage{natbib}       
\begin{document}

\newcommand{\ul}[1]{\underline{#1}}
\newcommand{\ol}[1]{\overline{#1}}
\newcommand{\ua}{\uparrow}
\newcommand{\ra}{\rightarrow}
\newcommand{\Ra}{\Rightarrow}
\newcommand{\Lra}{\Leftrightarrow}
\newcommand{\Llra}{\Longleftrightarrow}
\newcommand{\ds}{\displaystyle}

\begin{frontmatter}

\title{\large Colored Petri Nets for Modeling and Simulation of a Green Supply Chain System\thanksref{footnoteinfo}} 

\thanks[footnoteinfo]{The Erasmus+ International Credit Mobility (ICM) KA107 France-Indonesia 2020-2023 is acknowledged.}


\author[First,Second]{Daffa. R. Kaiyandra} 
\author[First]{F. Farizal} 
\author[Second]{Naly Rakoto}
%
\address[First]{Universitas Indonesia, Indonesia
    (e-mail: {\tt farizal@eng.ui.ac.id})}
\address[Second]{IMT Atlantique and LS2N, France, 
    (e-mail: {\tt naly.rakoto@imt-atlantique.fr})}

\begin{abstract}                
Green supply chain is an emerging approach in supply chain management to reduce environmental 
impact of the process concerning the flow of goods and materials. As a discrete-event system, supply 
chain   can be modeled using Petri Nets. Colored  Petri Nets (CPNs) extend the classical Petri net formalism 
with data, time and hierarchy. These extensions makes it possible to deal with the green aspects of the supply chain.

This paper deals with the Colored Petri Net approach to model and simulate a green supply chain system.
The forward supply chain network starts from the raw material suppliers, to the manufacturers, wholesalers, 
retailers and to the final customers. The reverse supply chain network is made of collecting points, recycling plants, 
disassembly plants,and secondary material market. A government environment agency plays the role of regulation 
of the recycling process.

A colored Petri net model of the  green supply chain is developed and simulated with CPN Tools, a 
dedicated software for CPN models. The simulation results as well as the state-space analysis results validate the correctness
of the model. 
\end{abstract}

\begin{keyword}
Green Supply Chain Management; Colored Petri Nets; Modeling; Simulation.
\end{keyword}

\end{frontmatter}

\section{Introduction}

Supply chain is a network of facilities and distribution options that performs the functions of procurement 
of materials, transformation of these materials into intermediate and finished products, and the distribution 
of these finished products to customers \citet{atasu-wassenhove12}. Supply chains are constantly upgraded 
to better meet changing  client needs and gain a competitive edge. Rising production and transportation costs 
required a more efficient  supply chain. Efficient supply chains are the ones that can deliver items to clients at a 
cheap cost while preserving product and service quality.
\vskip 0.3cm

Nowadays however, there are increasing awareness of the environmental impact of industrial activities, 
which leads companies aiming to reduce their carbon footprint, minimize waste, and use more sustainable 
material. Businesses are now also realizing that consumers are becoming more environmentally conscious 
and some are even willing to pay more for eco-friendly products. Governments around the world are also
 implementing strict environmental regulations to promote sustainable practices in industries and companies that 
 do not comply may face legal penalties which in turns may have becoming a significant loss of profit and opportunity. 
 \vskip 0.3cm
 
 The combination of these factors has led to a development of a new model of supply chain that takes into 
 account the environmental impact of each step, the new model is called green supply chain. Compared to 
 the traditional supply chain, green supply chain implies approaching factors according to the imperatives of
 the environment and adopting strategies in the sense of sustainable development. Adding the term “green” 
 to supply chain implies approaching the supply chain from the traditional perspective by corroboration/combination
 with the natural environment \cite{srivastava07}.
 \vskip 0.3cm
 
Green supply chain management has emerged as a widely acknowledged approach for reducing the 
environmental impact of processes concerning the flow of goods and materials throughout the supply 
chain. A green supply chain integrates eco-friendly elements into the traditional supply chain thereby contributing
 to the reduction of production costs, supporting economic growth and benefiting the environment ecologically. 
 The rising incline towards green supply chain in today’s industry is not only due to the growing ecological 
 concerns but also because of the competitive edge and economic advantages it offers when managed appropriately. 
 There are many aspects of green supply chain management, such as green design, green operations, waste 
 management, reverse logistics and network design, and green manufacturing and remanufacturing \cite{srivastava07}.
 \vskip 0.3cm

The supply chain, as well as the green supply chain, is a dynamic system that involves the constant flow 
of information, products, and money between different layers that are constantly changing. It also has a dynamic 
behavior due to changing customer demands and goals for competitive advantage that are desired to be achieved.
The ever-changing dynamics of the supply chain make modeling and analysis challenging.
\vskip 0.3cm

Many studies  \cite{zhang-lu-wu11} and \cite{fierro-cano-garcia20} defined supply chain network and green supply chain
\cite{ding-et-al.21} as discrete-event systems. Discrete events in the supply chain include: trucks departing from 
the supplier, materials arriving at the manufacturer, the manufacturing process beginning, trucks departing from the 
manufacturer to deliver finished products to the distributor, product delivered to the final customers \cite{zhang-lu-wu11}. 
\vskip 0.3cm

Petri Net, introduced by C. A. Petri (1962) is a classical tool for modeling, simulation, control and verification of discrete-event systems
\cite{cassandras21}, \cite{murata89}. See \cite{giua-silva18} for a historical perspective of Petri Nets in Control.
Colored Petri Nets (CPN)s have been introduced by K. Jensen (1992) in order to extend classical Petri Nets 
formalism with data, time and hierarchy. Many other extensions have been proposed e.g.  Hybrid Petri Nets (HPN)
\cite{david-alla.01} or  First-Order Hybrid Petri Nets (FOHPN) \cite{balduzzi-et-al.00}, \cite{dotoli-et-al.09},
\cite{balduzzi-et-al.10}.
\vskip 0.3cm

The paper is organized as follows. In Section 2, a colored Petri net model is given, and the methodology given in Section 3. 
In Section 4, simulation results are provided. Finally a conclusion is given in Section 5.

\section{Colored Petri Net Model}

 A colored Petri Net is a five-tuple $CPN = (P, T, Pre, Post,C)$ where:
\begin{itemize} 
\item $P$ is a set of places,
\item $T$ is a set of transitions,
\item $Pre: P \times T \ra N$ is the pre-incidence function, 
\item $Post: P \times T \ra N$ is the post-incidence function, 
\item $C: P \cup T \ra \{c_1, c_2, \dots c_n\}$ is the color function.
\end{itemize}

 \vskip 0.3cm

A common model of a green supply chain network has been developed by various authors 
\cite{srivastava07}, \cite{ding-et-al.21}, \cite{sheu-chou-hu05}, \cite{pan-wu14}.
Figure 1 is proposed as an example of the green supply chain model \cite{sheu-chou-hu05}. On this model there 
are common supply chain actors such as material suppliers, manufacturers, wholesalers, retailers, and 
end customers. Also, there are reverse supply chain network consisting of collecting points, recycling plants, 
disassembly plants, and secondary material market. There is also government environment agency who 
regulates the recycling process.
\vskip 0.3cm

On this green supply chain model, manufacturer can procure their supply from their usual raw material 
supplier or secondary material market which sells recycled materials. After they receive their supply, they 
can produce the products for the consumers to buy. If the wholesalers buy it, they can sell to retailers too. 
As for the end-customers, they have options on where to buy it. They can buy it directly from manufacturers, 
wholesalers, or retailers. Usually, customers can buy only from retailers, and if they want to buy the products 
in bulk or huge quantity, then they can buy from manufacturer or wholesalers.
\vskip 0.3cm

After the customers finishes using the products, they can send the products to collecting points, recycling 
plants, or directly to disassembly plants, depending on the condition of the used products. If they decided to 
send the used products, the customers will receive money. At the disassembly plants any reusable materials 
will be sold to secondary material market, where it would be sold again to the manufacturers. But, if the materials 
are in an unusable condition, they would be sent to the final disposal.
\vskip 0.3cm

On this model, manufacturers are obliged to pay a sum of recycling fee to the government environment 
agency where it would be used to subsidize disassembly plants. The reason for this is because in some 
countries, manufacturers are responsible for the waste of their products, so manufacturers have to have a 
processing plant for their used products. But, most of the manufacturers did not have the facility for that, so 
instead forcing the company to build a new processing plant, the government impose a regulation where 
manufacturers have to pay a recycling fee to the government environment agency or third-party environment 
organization, where the recycling process or waste processing will be done by a third- party entities such as 
recycling plant or disassembly plant.

\begin{table*}[t]
\caption{Petri Net Based Models of Supply Chains}
\begin{center}
\begin{tabular}{ |p{5cm}|p{6cm}|p{3cm}|}
 \hline
 \hline
 {\bf Authors}				& {\bf Scope}				& {\bf Methods} \\
 \hline
 \cite{dotoli-et-al.09}			& Forward Supply Chain		& FOHPN \\
 \hline
 \cite{van-der-aalst92}		& Forward Supply Chain		& CPN \\
 \hline
  \cite{srinivasa99}			& Forward Supply Chain		& Basic Petri Nets \\
 \hline
 \cite{makajic-et-al.04}		& Forward Supply Chain		& CPN \\
 \hline
 \cite{khilwani-et-al.10}		& Forward Supply Chain		& HPN \\
 \hline
 \cite{bevilacqua-et-al.12}		& Forward Supply Chain		& Timed CPN \\
 \hline
  \cite{mazzuto-et-al.12}		& Forward Supply Chain		& Timed CPN\\
 \hline
 \cite{santana12}			& Forward Supply Chain		&CPN \\
 \hline
 \cite{mukhlash18}			& Forward Supply Chain		& FOHPN \\
 \hline
 \cite{azougagh19}			& Forward Supply Chain		& CPN \\
 \hline
  \cite{idel-mahjoub21}		& Forward Supply Chain		& Timed CPN \\
 \hline
 \cite{hanafi-et-al.08}			& Reverse Supply Chain		& CPN \\
 \hline
 \cite{haiyan09}				& Reverse Supply Chain		& GSPN \\
 \hline
 \cite{murayama03}			& Forward and Reverse Supply Chain	&  CPN \\
 \hline
 \cite{wang13}				& Forward and Reverse Supply Chain	& Timed CPN \\
 \hline 
 \cite{pan-wu14} 			& Forward and Reverse Supply Chain	& GSPN \\
 \hline 
 \cite{outmal-et-al.16}		& Forward and Reverse Supply Chain	& FOHPN \\
 \hline 
 \cite{ding-et-al.21}			& Forward and Reverse Supply Chain	& GSPN \\
 \hline 
 \hline 
\end{tabular}
\end{center}
\end{table*}

\section{Methodology}

A Colored Petri Net of this model has been developed (see Figure 2 and Table 1 for the explanation) to 
analyze the properties of the system. The modelling is done in a tool for creating, modifying, verifying, and 
simulating CPN called  CPN Tools [33]. In the CPN model, the tokens are finite, it is to represent the 
properties of resources that are not unlimited. There are five color sets that represents five types of data: 
color set CASH to represent cash flow between the supply chain actors, color set MATERIAL to represent flow 
of material for the manufacturers, color set PRODUCTS to represent the finished products produced by the 
manufacturer, color set USED to represent used products or products that have reach its end-of-life, and lastly, 
color set WASTE to represent the flow of waste after the disassembly process.

\vskip 0.5cm
\begin{figure*}[t]
\begin{center}
\includegraphics[scale=0.25]{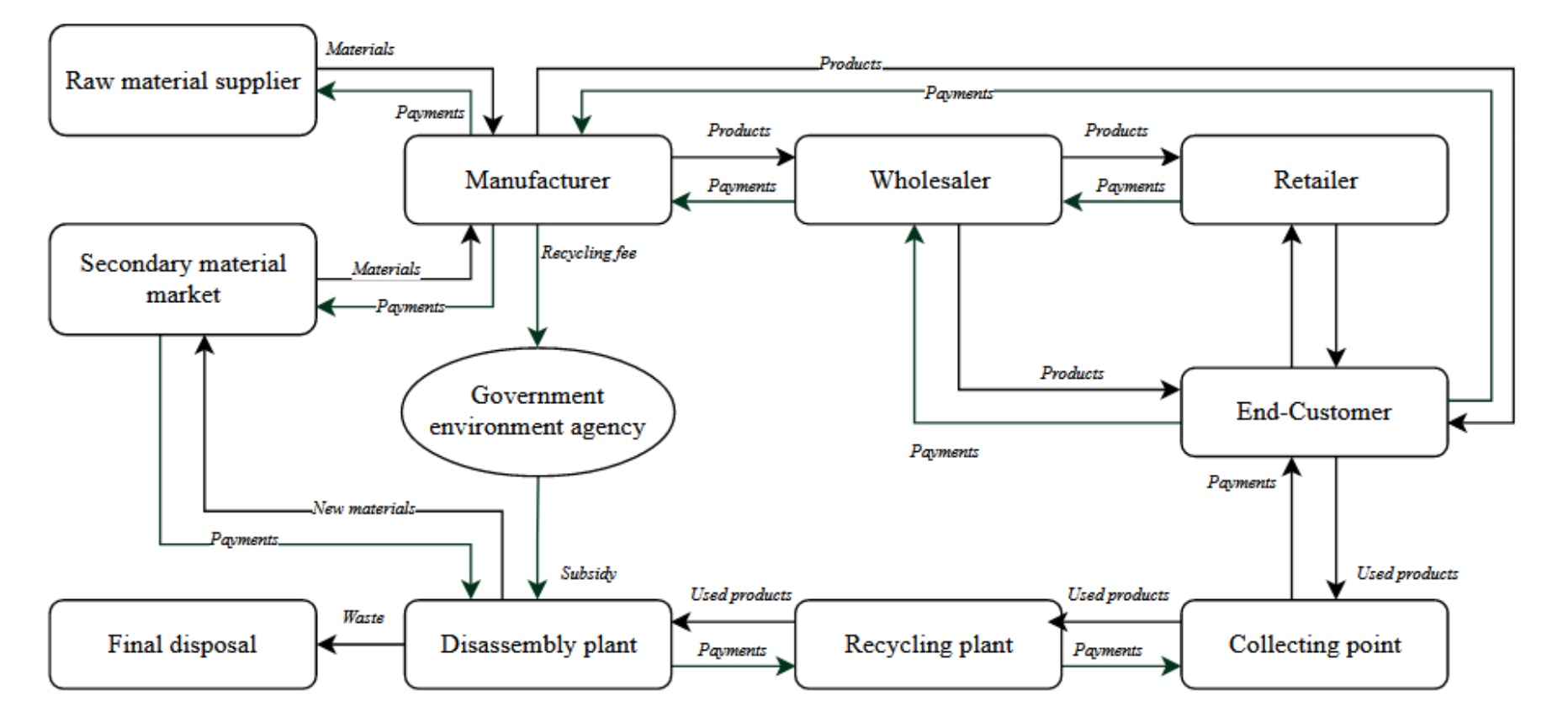}
\end{center}
\caption{\large Green Supply Chain Model (modified from \cite{sheu-chou-hu05})}
\label{green-scm1}
\end{figure*}
 \vskip 0.3cm

In the CPN model, there are two inhibitor arcs that connect P0’ (Manufacturer’s pre-production stage) and 
T1 (Manufacturer order material from raw material supplier) and T2 (Manufacturer buy material from secondary 
material market), the reason for this is to prevent manufacturer procure new materials when they already have 
the material. By doing this, we prevent the manufacturer to have large amount of inventory which in real-life will 
be costly for the manufacturing company. There is also an inhibitor arc to model a simple transaction, to enable 
an order transition (T1, T2, T3, T4, T5, T6, T7) the input transition of the place must have ‘cash’ from the actors 
(or places) who bought the items and ‘materials’ or ‘products’ from actors who sells it. The output will be ‘cash’ flow 
to the seller, and ‘materials’ or ‘products’ to the buyer.

\vskip 0.3cm

\begin{figure*}[h]
\begin{center}
\includegraphics[scale=0.25]{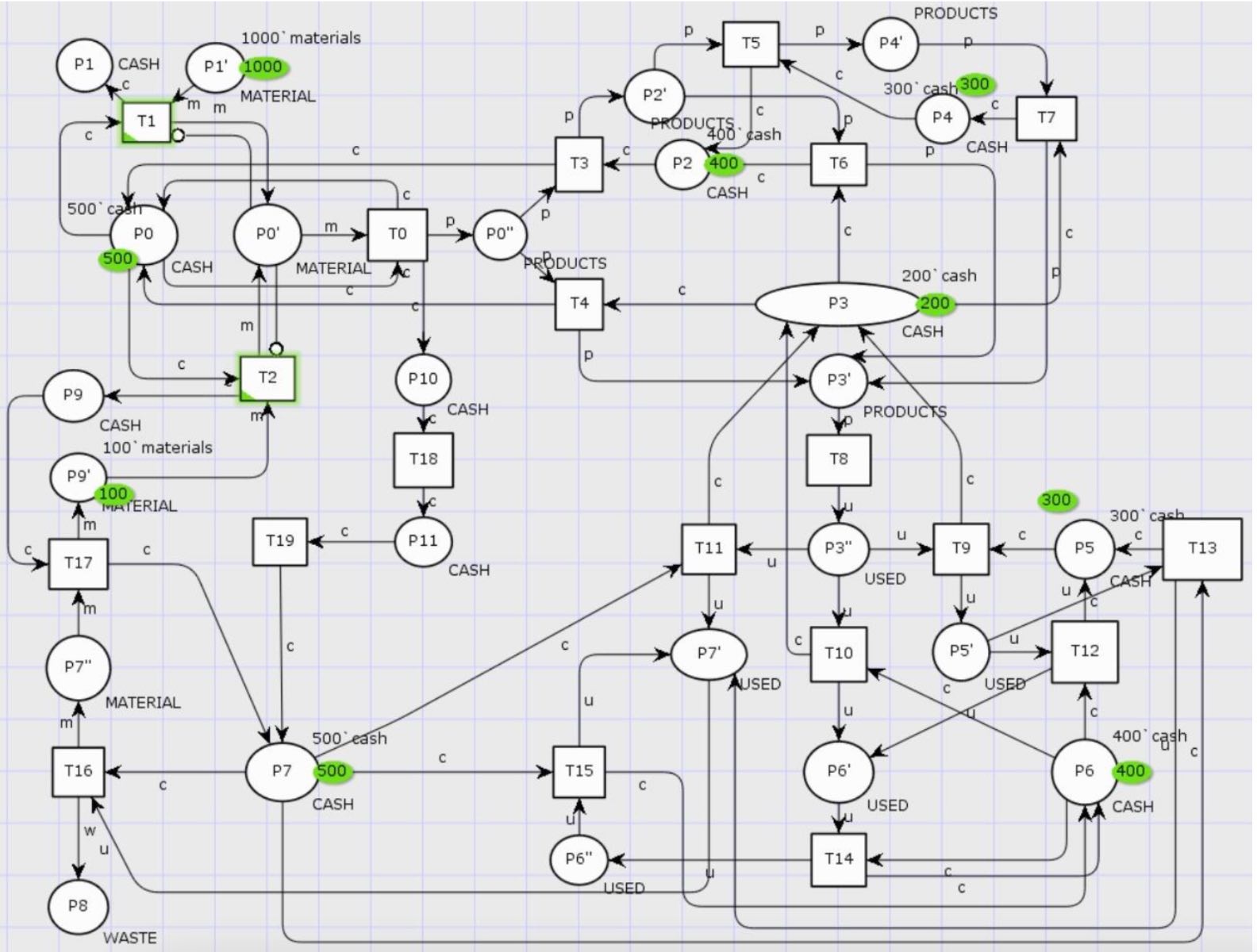}
\end{center}
\caption{\large Colored Petri Nets model of the green supply chain model}
\label{cpn2}
\end{figure*}
\vskip 0.2cm

For the latter half of the model, the reverse supply chain that starts from the end-customer, the process and 
also the place/transition modelling is similar but the real-life scenario is a little bit different. On the reverse supply 
chain, the end- customer can decide on where to send the products (on the CPN model, P3’’ have options to fire 
to T9, T10, or T11). In the green supply chain model, manufacturer have to pay a sum of fee to government 
environment agency who will subsidize the third-party disassembly processing plants. To model this, in every 
production process (or T0), manufacturer will send ‘cash’ to the government environment agency.

\begin{figure}[ht]
\begin{center}
\includegraphics[scale=0.9]{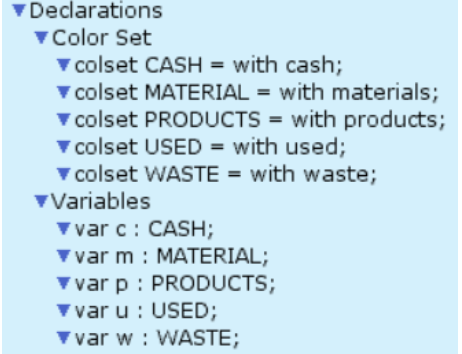}
\end{center}
\caption{\large Definition of color sets and variables}
\label{color-sets-var}
\end{figure}
~\vskip 0.5cm

\begin{center}
\begin{table*}
\caption{Description of Places and Transitions}
\begin{tabular}{ |p{1.2cm}|p{5.6cm}||p{1.2cm}|p{7.3cm}| }
 \hline
 \hline
 {\bf Places}& {\bf Description}					& {\bf \small Trans.} & {\bf Description}\\
 \hline
 {\bf P0}   	& Manufacturer's cashflow    			& {\bf T0}	& Production Process\\ 
 \hline
 {\bf P0'}  	& Manufacturer's pre-production stage    	& {\bf T1}	& Manufacturer orders material from raw mat. supplier\\ 
 \hline 
  {\bf P0''} 	& Manufacturer's post-production   		& {\bf T2}	& Manufacturer buys material from secondary mat. market\\  
 \hline
 {\bf P1}   	& Raw material supplier cash flow    		& {\bf T3}	& Wholesaler buys products from manufacturer\\ 
 \hline 
  {\bf P1'}   &  Raw material supplier warehouse  	& {\bf T4}	& Customer buys products from manufacturer \\ 
 \hline
 {\bf P2}   	&  Wholesaler's cashflow   			& {\bf T5}	& Retailer buys products from wholesaler\\ 
 \hline 
  {\bf P2'}  	&  Wholesaler with the product			& {\bf T6}	& Customer buys products from wholesaler \\  
 \hline
 {\bf P3}   	&  Customer's cashflow   				& {\bf T7}	& Customer buys products from retailer\\ 
 \hline 
 {\bf P3'}   	&  Customer with the product   			& {\bf T8}	& Customer uses the products \\ 
 \hline
 {\bf P3''}  	&  Customer with the used/finished product   & {\bf T9}& Customer sends used products to collecting point \\ 
 \hline 
  {\bf P4} 	&   Retailer's cashflow 				& {\bf T10}  	& Customer sends used products to recycling plant \\  
 \hline
 {\bf P4'}   	&  Retailer with the product    			& {\bf T11} 	& Customer sends used products to disassembly plant \\ 
 \hline 
  {\bf P5}   &  Collecting point cashflow 			& {\bf T12}	& Collection point sends used products to recycling plant \\ 
 \hline
 {\bf P5'}   	&   Collecting point with the used products	 & {\bf T13}	& Collection point sends used products to disassembly pts \\ 
 \hline 
  {\bf P6}  	&  Recycling plant cashflow			& {\bf T14}	& Recycling point recycles the used products \\  
 \hline
 {\bf P6'}   &   Used products waiting to be recycled & {\bf T15}	&  Recycling point sends used products to disassembly plant  \\ 
  \hline
 {\bf P6''}   &  Used products after recycling process  & {\bf T16}	& Disassembly process  \\ 
 \hline
 {\bf P7}  	&  Disassembly plant cashflow 		& {\bf T17} & Disassembly plant sells materials to second. mat. market\\ 
 \hline 
  {\bf P7'} 	&  Disassembly plant (used / recycled prod.) & {\bf T18}&  Manufacturer pays fee to govt environment agency  \\
 \hline
 {\bf P7''}   & New materials obtained after disassembly     & {\bf T19} &  Govt envt. agency subsidizes the disassembly plants\\ 
 \hline 
  {\bf P8}   &  Final disposal for waste 			& ~			&  ~ \\ 
 \hline
 {\bf P9}   	&  Secondary supply market cashflow 	& ~			&  ~ \\ 
 \hline 
  {\bf P9'}  	&  Secondary supply market warehouse	& ~ 			& ~ \\  
 \hline
 {\bf P10}   &  Manufacturer's mandatory recycling fee & ~ 		&  ~ \\ 
 \hline
 {\bf P11}   &   Government environment agency	& ~ 			&  ~ \\ 
 \hline
 \hline
\end{tabular}
\end{table*}
\end{center}

\section{Simulation Results}

The green supply chain has been simulated with the CPN Tools software.  50.000 random firings have 
been performed. The results of the simulation are presented in Table 2. From Table 2, the number of
firing of each transition is given. Transition T0 (production process by the manufacturer) is the transition 
that has been fired the most (4703 times).1000 times come from raw material supplier, while 3703 times
come from secondary material market.
Then come the 3 transitions T8, T18 and T19 which fires 4702 times each. These high numbers of firings
is explained by the fact that transition T18 and T19 receive tokens from place P10, which receives tokens
every time the production process is active. Transition T8 has a high number of firings because its input place
(P3') receives tokens from both T4 and T7.Transition T1 only fired 1000 times because place P1' has a maximum 
capacity of 1000 tokens. Customers mostly buy the products directly from the manufacturer (T4, 2344 times)
 more than from the wholesaler (T6, 1176 times) or from the retailer (T7, 1183 times).     
 \vskip 0.3cm
 
\begin{figure}[ht]
\begin{center}
\includegraphics[scale=0.5]{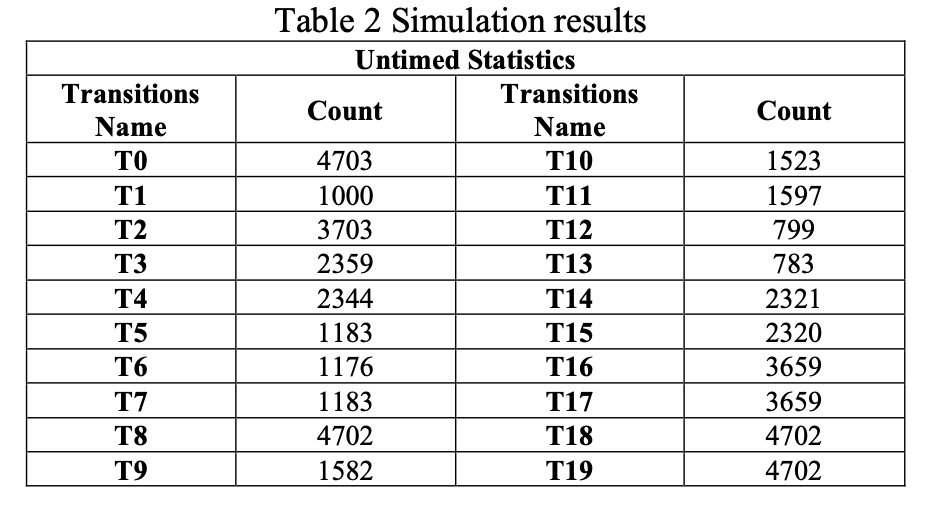}
\end{center}
\label{sim-res1}
\end{figure}
\vskip 0.2cm

For the reverse supply chain, it seems that customers are balanced between sending the used products 
to the collecting points (T9, 1582 times), recycling plant (T10, 1523 times), or disassembly plant
(T11, 1597 times). Finally  the flow of products from collecting  points to the disassembly plant (T13) and to 
the recycling plant (T12) have the lowest number of firings.
\vskip 0.2cm

The state-space analysis results are as follows. There are 20.252 possible  states  with 7.237 possible 
dead markings on this model and 86.533 arcs that are connecting each and every 
states. The SCC (Strongly Connected Components) graph also has the same results which means there are 
no cycles or subgraph on this model. Also, after fairness analysis, it is found that there are no infinite occurrence 
sequences, which means eventually there will be no transitions available to be fired.
 \vskip 0.2cm
 
\begin{figure}[ht]
\begin{center}
\includegraphics[scale=0.55]{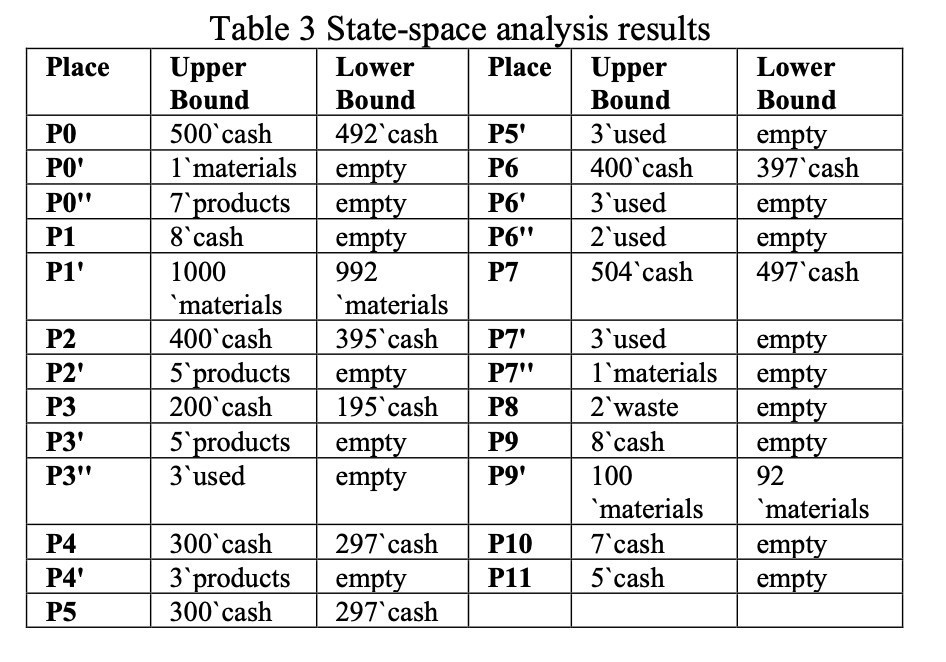}
\end{center}
\label{state-space-res2}
\end{figure}
\vskip 0.2cm

There is no home marking on this CPN, home marking is a marking which can be reached from any 
reachable marking, if there are no home marking, then it means that all markings are different each time. 
 On our CPN model, all the places are bounded. While the 
lower bound explains the minimal number of tokens on the place in a reachable marking. Few of our places have 
lower bound of 0, this can be explained as most of them have 0 token in the initial marking.

\section{Conclusions}
This paper has presented a Colored Petri Net approach to model and simulate a green supply chain system.
 A colored Petri net model of the  green supply chain is developed and simulated with CPN Tools \cite{cpn-tools}, a 
dedicated software for CPN models. The simulation results as well as the state-space analysis results validate the correctness
of the model. Future work will deal with the generalisation of the method to a class of Green Supply Chain Systems. 
Another future work will focus on the Reverse Supply Chain part and the strategies dedicated to it.


\begin{thebibliography}{xx}  	

\bibitem[Atasu and Van Wassenhove (2012)]{atasu-wassenhove12}
A. Atasu and L.N. Van Wassenhove (2012).
\newblock "An operations perspective on product take-back legislation for E-waste: Theory, practice, 
and research needs,"
\newblock In \emph {Prod. Oper. Manag.}, vol. 21, no. 3, pp. 407– 422.

\bibitem[Azougagh et~al.(2019)]{azougagh19}
Y. Azougagh, K. Benhida, S. Elfzazi (2019).
\newblock "Modeling Supply Chains Using Colored Petri Nets: Application in a Phosphate Supply Chain" 
\newblock In \emph{J. Mech. Cont. and Math. Sci.}, Special Issue, N. 4 (2019) pp. 266-276.

\bibitem[Balduzzi et~al.(2000)]{balduzzi-et-al.00}
F. Balduzzi, A. Giua, and G. Menga (2000).
\newblock "First-Order Hybrid Petri Nets: A Model for Optimization and Control",
\newblock In \emph{IEEE Transactions on Robotics and Automation}, vol 16, N.4, pp-382--399.

\bibitem[Balduzzi et~al.(2010)]{balduzzi-et-al.10}
F. Balduzzi, A. Giua, and C. Seatzu (2010).
\newblock "Modelling and Simulation of Manufacturing Systems with First-Order Hybrid Petri Netsl",
\newblock In \emph{Int. Journal of Production Research}, vol 39, N.2, pp-255--282.

\bibitem[Bevilacqua et~al.(2012)]{bevilacqua-et-al.12}
M. Bevilacqua, F.E. Ciarapica, and G. Mazzuto (2012).
\newblock "Modelling and performance analysis of a supply chain using timed coloured Petri nets,"
\newblock  In \emph{Int. J. of Business Performance and Supply Chain Modelling, IJBPSCM} Vol. 4 , No. 3/4, pp. 285--316.
 
\bibitem[Cassandras and Lafortune (2021)]{cassandras21}
C.G. Cassandras and S. Lafortune (2021).
\newblock "Introduction to Discrete Event Systems,"
\newblock 2nd Edition, Springer Verlag.

\bibitem[David and Alla (2001)]{david-alla.01}
R. David and H. Alla (2001).
\newblock "On Hybrid Petri Nets",
\newblock In \emph{Discrete Event Dynamic Systems}, vol 11 (2001), pp-9--40.

\bibitem[Ding et~al. (2021)]{ding-et-al.21}
J. Ding, X. Chen, H. sun, W. Yan, and H. Fang (2021).
\newblock "Hierarchical Structure of a Green Supply Chain"
\newblock In \emph{Computers and Industrial Engineering}, vol 157 (2021),  pp. 1--10.

\bibitem[Dolgui et~al. (2018)]{dolgui-et-al.18}
A. Dolgui, D. Ivanov, S. Sethi, and B. Sokolov (2018).
\newblock "Control Theory Applications To Operations Systems, Supply Chain Management and Industry 4.0 Networks,"
\newblock In \emph{IFAC-PapersOnLine} vol 51 No 11, pp. 1536--1541.

\bibitem[Dotoli et~al.(2009)]{dotoli-et-al.09}
M. Dotoli, M.P. Fanti, G. Iacobellis, and A.M. Mangini (2009).
\newblock "A First-Order Hybrid Petri Net Model for Supply Chain Management",
\newblock In \emph{IEEE Transactions on Automation Science and Engineering}, vol 6, N.4, pp-744-758.

\bibitem[Fierro et~al. (2020)]{fierro-cano-garcia20}
L.H. Fierro, R.E. Cano, and J.I. Garcia (2020).
\newblock "Modelling of a Multi-Agent Supply Chain Management System using Colored Petri Nets"
\newblock In \emph{Procedia Manufacturing}, vol 42 (2020),  pp. 288--295.

\bibitem[Guia and Silva (2018)]{giua-silva18}
A. Giua and M. Silva (2018).
\newblock "Petri Nets and Automatic Control: A Historical Perspective"
\newblock In \emph{Annual Review in Control},vol 45,  pp. 223--239.

\bibitem[Hafilah et~al. (2019)]{HCLR19}
D.L. Hafilah, A. Cakravastia, Y. Lafdail, and N. Rakoto (2019).
\newblock "Modeling and Simulation of Air France Baggage Handling System with Colored Petri Nets,"
\newblock Proc. of IFAC MIM 2019, IFAC-PapersOnLine 52-13 (2019) pp. 2443--2448. 

\bibitem[Hanafi (2008)]{hanafi08}
J. Hanafi (2008).
\newblock "Modeling of Collection Strategies for End-of-Life Products using Petri Net,"
\newblock PhD Thesis, The Univ. of New South Wales, Australia.

\bibitem[Hanafi et~al. (2008)]{hanafi-et-al.08}
J. Hanafi, S. Kara, and H. Kaebernick (2008).
\newblock Reverse Logistics Strategies for End-of-Life Products.
\newblock In \emph{Int. J. of Logistics Management} vol 19 N. 3, pp. 367--388.

\bibitem[Idel Mahjoub (2021)]{idel-mahjoub21}
Y. Idel Mahjoub, E. H. Chakir, and A. Nait-Sidi-Moh (2021).
\newblock "Logistic Networl Modeling and Optimization: An Approach based on (max,+) Algebra and Coloured  Petri Nets,"
\newblock In \emph{Computers and Industrial Engineering},vol 158 (2021) 107341.

\bibitem[Jensen (1992)]{jensen92}
K. Jensen (1992).
\newblock "Coloured Petri Nets: Basic Concepts, Analysis Methods and Practical Use," Vol. 1: Basic Concepts
\newblock Springer Verlag.

\bibitem[Jensen and Kristensen (2009)]{jensen-kristensen09}
K. Jensen and L. Kristensen (2009).
\newblock "Coloured Petri Nets: Modeling and Validation of Concurrent Systems,"
\newblock Springer Verlag, New York.

\bibitem[Kaiyandra et~al. (2023)]{kaiyandra-et-al.23}
D.R. Kaiyandra, F. Farizal, and N. Rakoto (2023).
\newblock "Petri Nets Application for Supply Chain Management: A Review of Recent Literature,"
\newblock Proc. of IEEE CoDIT 2023, Rome, Italy, pp. 1391--1396.

\bibitem[Khilwani et~al. (2010)]{khilwani-et-al.10}
N. Khilwani, M.K. Tiwari, and I. Sabuncuoglu (2010).
\newblock "Hybrid Petri Nets for Modelling and Performance Evaluation of Supply Chains'"
\newblock In \emph{Int. J. of Production Research} vol 49 N. 15, pp. 4627--4656.

\bibitem[Makajic-Nikolic et~al. (2004)]{makajic-et-al.04}
D. Makaji-Nikolic, B. Panic, M. Vujosevic (2004).
\newblock "Bullwhip Effect and Supply Chain Modelling and Analysis using CPN Tools.
\newblock In \emph{5th Workshop and Tutorial on Practical Use of Coloured Petri Nets and the CPN Tools, Aarhus, Denmark}, pp. 219--234.

\bibitem[Malcon and Martinez-Florez (2012)]{malcon-martinez12}
C. Malcon Cervera and J.L. Martinez Flores (2012).
\newblock "A Conceptual Model for a Green Supply Chain Strategy,"
\newblock In \emph{Glob. Conf. Bus. Financ. Proc.}, vol. 7, no. 2, pp. 269–274.

\bibitem[Mazzuto et~al. (2012)]{mazzuto-et-al.12}
G. Mazzuto, M. Bevilacqua, and F.E. Ciarapica (2012).
\newblock "Supply chain modelling and managing, using timed coloured Petri nets: a case study,"
\newblock In \emph{Int. J. of Production Research} vol 50 N. 16, pp. 4718--4733.

\bibitem[Mukhlash et~al. (2018)]{mukhlash18}
\newblock I. Mukhlash, W.N. Rumana, D. Adzkiya, and R. Sarno  (2018).
\newblock "Business Process Improvement of Production Systems using Coloured Petri Nets"
\newblock In \emph{Bulletin of Electrical Engineering and Informatics}, vol.7, N. 1 (2018),  pp. 102--112.

\bibitem[Murata (1989)]{murata89}
T. Murata (1989).
\newblock "Petri Nets: Properties, Analysis and Applications"
\newblock In \emph{Proceedings of the IEEE}, 77(4) pp. 541--580.

\bibitem[Murayama et~al. (2003)]{murayama03}
T. Murayama, S. Hatakenaka, and F. Oba (2003).
\newblock "Simulation-Based Evaluation of Forward and Reverse Supply Chain,"
\newblock Proc. of 3rd Int. Symp. on Environmentally Conscious Design and Inverse Manufacturing,  
2003 EcoDesign, Tokyo, Japan, pp. 521--526.

\bibitem[Nakano (2020)]{nakano20}
M. Nakano (2020)
\newblock "Supply Chain Management: Strategy and Organization,"
\newblock Springer Singapore.

\bibitem[Outmal et~al. (2016)]{outmal-et-al.16}
I. Outmal, A. Kamrani, E.S. Abouel Nasr, and M. Alkahtani (2016).
\newblock "Modeling and Performance Analysis of a Closed-Loop Supply Chain Using First-Order Hybrid Petri Nets,"
\newblock In \emph{Advances in Mechanical Engineering}, vol. 8 no. 5, pp. 1–15

\bibitem[Pan and Wu (2014)]{pan-wu14}
M. Pan and W. Wu (2014).
\newblock "A Petri Net Approach for Green Supply Chain Network for Modeling and Performance Analysis"
\newblock In \emph{N. Lohmann, M. Song and P. Wohed  (Eds), BPM 2013, Lecture Notes in Business Inf. Processing},
vol 171, Springer, Cham., pp 330-341.

\bibitem[Petri (1962)]{petri62}
\newblock C.A. Petri (1962).
\newblock "Communication with Automata."
\newblock PhD Thesis, Technische Universitaet Darmstadt, Germany.

\bibitem[Santana-Robles et~al.( 2012)]{santana12}
F. Santana-Robles, J. Medina-Martin, O. Montano-Arango, J.C. Seck-Tuoh-Mora (2012).
\newblock "Modeling and Simulation of Textile Supply Chain through Colored Petri Nets.
\newblock In \emph{Intelligent Information Management}, vol. 02 (25) pp. 261--268.

\bibitem[Sheu et~al. (2005)]{sheu-chou-hu05}
J.B. Sheu, Y.H. Chou, and C.C. Hu (2005).
\newblock "An integrated logistics operational model for green-supply chain management,"
\newblock In \emph{Transp. Res. Part E Logist. Transp. Rev.} ,vol. 41, no. 4, pp. 287–313.

\bibitem[Srinivasa and Viswanadham (1999)]{srinivasa99}
N.R. Srinivasa Raghavan and N. Viswanadham (1999).
\newblock "Performance Analysis of Supply Chain Networks using  Petri Nets,"
\newblock Proc. of the 38th IEEE CDC,  Phoenix, AZ, USA, pp. 57--62.

\bibitem[Srivastava (2007)]{srivastava07}
S.K. Srivastava (2007).
\newblock "Green supply chain management: A state-of- the-art literature review,"
\newblock In \emph{Int. J. Manag. Rev.}, vol. 9, no. 1, pp. 53–80.

\bibitem[van der Aalst (1992)]{van-der-aalst92}
W.M.P. van der Aalst (1992).
\newblock "Timed Coloured Petri Nets and their Applications to Logistics,"
\newblock PhD Thesis, T.U. Eindhoven, The Netherlands.

\bibitem[Wang (2013)]{wang13}
J.Y. Wang (2013).
\newblock "Modeling closed loop supply chain system based on stochastic fuzzy Petri net.
\newblock Proc. of the 20th Int. Conf. on Management Science and Engineering, Harbin, P.R. China,, pp. 598--603.

\bibitem[Ye et~al. (2020)]{ye-et-al.20}
X. Ye, D Ge, X. Bian, Q. Xu, and Y. Zhou (2020).
\newblock "Improving Business Process Efficiency for Supply Chain Finance: Empirical Analysis and Optimization
Based on Stochastic Petri Net,"
\newblock In \emph{IEEE Access}, D.O.I. 10.1109/ACCESS.2020.2995851

\bibitem[Zhang et~al. (2011)]{zhang-lu-wu11}
X. Zhang, Q. Lu, and T. Wu (2011).
\newblock "Petri-net based applications for supply chain management: An overview,"
\newblock In \emph{Int. J. Prod. Res.}, vol. 49, no. 13, pp. 3939–3961.

\bibitem[Haiyan and Shengqiang (2009)]{haiyan09}
Z. Haiyan and L. Shengqiang (2009)
\newblock ""Modeling and Analysis of Reverse Supply Chain Based on Generalized Stochastic Petri Nets,"
\newblock Proc. of 2009 INt. Conf. on Information Management, Innovation Management and Industrial Engineering,
Xi'an, China, pp. 437--440.

\bibitem[CPN Tools (2024)]{cpn-tools}
Colored Petri Nets Tools (Software),
\newblock "CPN Tools"
\newblock http://cpntools.org.

\end{thebibliography}

\end{document}